\documentclass[12pt]{article}
\newcommand{\sect}[1]{\setcounter{equation}{0}\section{#1}}

\renewcommand{\appendix}{\setcounter{section}{0}
\renewcommand{\thesection}{\Alph{section}}}

\def\a{\alpha}
\def\b{\beta}
\def\g{\gamma}
\def\G{\Gamma}
\def\d{\delta}

\def\e{\epsilon}

\def\k{\kappa}
\def\l{\lambda}
\def\m{\mu}
\def\n{\nu}
\def\r{\rho}
\def\s{\sigma}
\def\S{\Sigma}

\def\be{\begin{equation}}
\def\ee{\end{equation}}
\def\ba{\begin{eqnarray}}
\def\ea{\end{eqnarray}}

\newcommand{\nn}{\nonumber\\}
\newcommand{\no}{\nonumber}
\begin{document}
\renewcommand{\thefootnote}{\fnsymbol{footnote}}

\newpage
\setcounter{page}{0}
\pagestyle{empty}

\begin{center}
{\Large{\bf Semi-simple $o(N)$-extended super-Poincar\'e algebra\\}}
\vspace{1cm}
{\bf Dmitrij V. Soroka\footnote{E-mail: dsoroka@kipt.kharkov.ua} and 
Vyacheslav A. Soroka\footnote{E-mail: vsoroka@kipt.kharkov.ua}}
\vspace{1cm}\\
{\it Kharkov Institute of Physics and Technology,\\
1, Akademicheskaya St., 61108 Kharkov, Ukraine}\\
\vspace{1.5cm}
\end{center}
\begin{abstract}
A semi-simple tensor extension of the Poincar\'e algebra is 
given for the arbitrary dimensions $D$. It is illustrated that this 
extension is a direct sum of the $D$-dimensional Lorentz algebra $so(D-1,1)$ 
and $D$-dimensional anti-de Sitter (AdS) algebra $so(D-1,2)$. A supersymmetric 
also semi-simple $o(N)$ generalization of this extension is introduced in 
the $D=4$ dimensions. It is established that this generalization is a direct 
sum of the 4-dimensional Lorentz algebra $so(3,1)$ and orthosymplectic algebra 
$osp(N,4)$ (super-AdS algebra). Quadratic Casimir operators for the 
generalization are constructed. The form of these operators indicates that 
the components of an irreducible representation for this generalization are 
distinguished by the mass, angular momentum and quantum numbers corresponding 
to the internal symmetry, tensor and supersymmetry generators. That generalizes
the Regge trajectory idea.

The probable unification of the $N=10$ supergravity with the $SO(10)$ GUT 
model is discussed.

This paper is dedicated to the memory of Anna Yakovlevna Gelyukh.

\bigskip
\noindent
{\it PACS:} 02.20.Sv; 11.30.Cp; 11.30.Pb; 11.15.-q

\medskip
\noindent
{\it Keywords:} Poincar\'e algebra, Tensor, Extension, Casimir operators,
Supersymmetry, Gauge group

\end{abstract}

\newpage
\pagestyle{plain}
\renewcommand{\thefootnote}{\arabic{footnote}}
\setcounter{footnote}0

\sect{Introduction}

In the papers \cite{bcr,sch,bh,gios1,gios2,cj,can,ss1,dss0,dss,ss2,ss3,
sorsor,bg,bg1,gkl,ggp,bgkl,l} the Poincar\'e algebra for the generators of
the rotations $M_{ab}$ and translations $P_a$ in $D$ dimensions
\begin{eqnarray}
[M_{ab},M_{cd}]=(g_{ad}M_{bc}+g_{bc}M_{ad})-(c\leftrightarrow d),\nonumber
\end{eqnarray}
\begin{eqnarray}
[M_{ab},P_c]=g_{bc}P_a-g_{ac}P_b,\nonumber
\end{eqnarray}
\begin{eqnarray}\label{1.1}
[P_a,P_b]=0
\end{eqnarray}
has been extended by means of the second rank tensor generator $Z_{ab}$ in
the following way:
\begin{eqnarray}
[M_{ab},M_{cd}]=(g_{ad}M_{bc}+g_{bc}M_{ad})-(c\leftrightarrow d),\nonumber
\end{eqnarray}
\begin{eqnarray}
[M_{ab},P_c]=g_{bc}P_a-g_{ac}P_b,\nonumber
\end{eqnarray}
\begin{eqnarray}
[P_a,P_b]=cZ_{ab},\nonumber
\end{eqnarray}
\begin{eqnarray}
[M_{ab},Z_{cd}]=(g_{ad}Z_{bc}+g_{bc}Z_{ad})-(c\leftrightarrow d),\nonumber
\end{eqnarray}
\begin{eqnarray}
[P_a,Z_{bc}]=0,\nonumber
\end{eqnarray}
\begin{eqnarray}\label{1.2}
[Z_{ab},Z_{cd}]=0,
\end{eqnarray}
where $c$ is some constant\footnote{Note that, the summation
over every pair of the antisymmetric indices is carried out with the factor 
${1\over2}$.}.

Such an extension makes common sense, since it is homomorphic to the usual 
Poincar\'e  algebra (\ref{1.1}). Moreover, in the limit ${c\to 0}$ the algebra
(\ref{1.2}) goes to the semi-direct sum of the commutative ideal $Z_{ab}$ and
Poincar\'e algebra (\ref{1.1}).

It is remarkable
enough that the momentum square Casimir operator of the Poincar\'e algebra
under this extension ceases to be the Casimir operator and it is generalized
by adding the term linearly dependent on the angular momentum
\begin{eqnarray}\label{1.3}
P^aP_a+cZ^{ab}M_{ba}\mathrel{\mathop=^{\rm def}}X_kh^{kl}X_l,
\end{eqnarray}
where $X_k=\{P_a, Z_{ab}, M_{ab}\}$. Due to this fact, an irreducible
representation of the extended algebra (\ref{1.2}) has to contain the fields 
with the different masses \cite{ss1,ss2,ss3}.
This extension with non-commuting momenta has also something in common
with the ideas of the papers~\cite{sn,ya,hl} and with the non-commutative
geometry idea~\cite{c}.

It is interesting to note that in spite of the fact that the algebra 
(\ref{1.2}) is not semi-simple and therefore has a degenerate Cartan-Killing
metric tensor nevertheless there exists another non-degenerate invariant
tensor $h_{kl}$ in the adjoint representation which corresponds to the
quadratic Casimir operator (\ref{1.3}), where the matrix $h^{kl}$ is inverse
to the matrix $h_{kl}$: $h^{kl}h_{lm}=\d_m^k$.

There are other quadratic Casimir operators
\ba\label{1.4}
c^2Z^{ab}Z_{ab},
\ea
\ba\label{1.5}
c^2\e^{abcd}Z_{ab}Z_{cd}.
\ea
Note that the Casimir operator (\ref{1.5}), dependent on the Levi-Civita
tensor $\e^{abcd}$, is suitable only for the $D=4$ dimensions.

It has also been shown that for the dimensions $D=2,3,4$ the extended
Poincar\'e algebra (\ref{1.2}) allows the following supersymmetric
generalization:
\begin{eqnarray}
\{Q_\a,Q_\b\}=-d(\sigma^{ab}C)_{\a\b}Z_{ab},\nonumber
\end{eqnarray}
\begin{eqnarray}
[M_{ab},Q_\a]=-(\sigma_{ab}Q)_\a,\nonumber
\end{eqnarray}
\begin{eqnarray}
[P_a,Q_\a]=0,\nonumber
\end{eqnarray}
\begin{eqnarray}\label{1.6}
[Z_{ab},Q_\a]=0
\end{eqnarray}
with the help of the super-translation generators $Q_\a$. In (\ref{1.6}) $C$ 
is a charge conjugation matrix, $d$ is some constant and 
$\sigma_{ab}={1\over4}[\gamma_a,\gamma_b]$, where $\g_a$ is the Dirac matrix.
Under this supersymmetric generalization the quadratic Casimir operator
(\ref{1.3}) is modified into the following form:
\begin{eqnarray}\label{1.7}
P^aP_a+cZ^{ab}M_{ba}-{c\over2d}Q_\a(C^{-1})^{\a\b}Q_\b,
\end{eqnarray}
while the form of the rest quadratic Casimir operators (\ref{1.4}), 
(\ref{1.5}) remains unchanged.

In the present paper we propose another possible semi-simple tensor extension 
of the $D$-dimensional Poincar\'e algebra (\ref{1.1}) which turns out a direct
sum of the $D$-dimensional Lorentz algebra $so(D-1,1)$ and $D$-dimensional 
anti-de Sitter (AdS) algebra $so(D-1,2)$. For the case $D=4$ dimensions we 
give for this extension a supersymmetric $o(N)$ generalization which is a 
direct sum 
of the 4-dimensional Lorentz algebra $so(3,1)$ and orthosymplectic algebra 
$osp(N,4)$ (super-AdS algebra). In the limit this 
supersymmetrically generalized extension go to the Lie superalgebra
(\ref{1.2}), (\ref{1.6}).

Let us note that the introduction of the semi-simple $o(N)$-extended 
super-Poincar\'e algebra is very important for the construction of the models, 
since it is easier to deal with the non-degenerate space-time symmetry.

\sect{Semi-simple tensor extension}

In the paper \cite{sorsor} we extended the Poincar\'e algebra 
(\ref{1.1}) in the $D$ dimensions by 
means of the tensor generator $Z_{ab}$ in the following way:
\begin{eqnarray}
[M_{ab},M_{cd}]=(g_{ad}M_{bc}+g_{bc}M_{ad})-(c\leftrightarrow d),\nonumber
\end{eqnarray}
\begin{eqnarray}
[M_{ab},P_c]=g_{bc}P_a-g_{ac}P_b,\nonumber
\end{eqnarray}
\begin{eqnarray}
[P_a,P_b]=cZ_{ab},\nonumber
\end{eqnarray}
\begin{eqnarray}
[M_{ab},Z_{cd}]=(g_{ad}Z_{bc}+g_{bc}Z_{ad})-(c\leftrightarrow d),\nonumber
\end{eqnarray}
\begin{eqnarray}
[Z_{ab},P_c]={4a^2\over c}(g_{bc}P_a-g_{ac}P_b),\nonumber
\end{eqnarray}
\begin{eqnarray}\label{2.1}
[Z_{ab},Z_{cd}]={4a^2\over c}[(g_{ad}Z_{bc}+g_{bc}Z_{ad})
-(c\leftrightarrow d)],
\end{eqnarray}
where $a$ and $c$ are some constants. This Lie algebra, when the quantities
$P_a$ and $Z_{ab}$ are taken as the generators of a homomorphism kernel, is  
homomorphic to the usual Lorentz algebra.  It is remarkable that the Lie 
algebra (\ref{2.1}) is {\it semi-simple} in contrast to the Poincar\'e
algebra (\ref{1.1}) and extended Poincar\'e algebra (\ref{1.2}).

The extended Lie algebra (\ref{2.1}) has the following quadratic Casimir
operators:
\begin{eqnarray}\label{2.2}
C_1=P^aP_a+cZ^{ab}M_{ba}+2a^2M^{ab}M_{ab}\mathrel{\mathop=^{\rm def}}
X_kH_1^{kl}X_l,
\end{eqnarray}
\ba\label{2.3}
C_2=c^2Z^{ab}Z_{ab}+8a^2(cZ^{ab}M_{ba}+2a^2M^{ab}M_{ab})\mathrel{\mathop=^{\rm def}}
X_kH_2^{kl}X_l,
\ea
\ba\label{2.4}
C_3=\e^{abcd}[c^2Z_{ab}Z_{cd}+8a^2(cZ_{ba}M_{cd}+2a^2M_{ab}M_{cd})].
\ea
Note that in the limit $a\to0$ the algebra (\ref{2.1}) tend to the algebra
(\ref{1.2}) and the quadratic Casimir operators (\ref{2.2}), (\ref{2.3}) and
(\ref{2.4}) are turned into (\ref{1.3}), (\ref{1.4}) and (\ref{1.5}),
respectively.

The symmetric tensor
\ba\label{2.5}
H^{kl}=sH_1^{kl}+tH_2^{kl}=H^{lk}
\ea
with arbitrary constants $s$ and $t$ is invariant with respect to the adjoint
representation
\ba
H^{kl}=H^{mn}{U_m}^k{U_n}^l.\no
\ea
Conversely, if we demand the invariance with respect to the adjoint representation
of the second rank contravariant symmetric tensor, then we come to the structure
(\ref{2.5}) (see also the relation (32) in \cite{dss}).

The semi-simple algebra (\ref{2.1})
\ba
[X_k,X_l]={f_{kl}}^mX_m\no
\ea
has the non-degenerate Cartan-Killing metric tensor
\ba
g_{kl}={f_{km}}^n{f_{ln}}^m,\no
\ea
which is invariant with respect to the co-adjoint representation
\ba
g_{kl}={U_k}^m{U_l}^ng_{mn}.\no
\ea
With the help of the inverse metric tensor $g^{kl}$: $g^{kl}g_{lm}=\d_m^k$ we 
can
construct the quadratic Casimir operator which, as it turned out, has the
following expression in terms of the quadratic Casimir operators (\ref{2.2})
and (\ref{2.3}):
\ba\label{2.6}
X_kg^{kl}X_l={1\over8a^2(D-1)}\left[C_1+{3-2D\over8a^2(D-2)}C_2\right],
\ea
that corresponds to the particular choice of the constants $s$ and $t$ in
(\ref{2.5}).

The extended Poincar\'e algebra (\ref{2.1}) can be rewritten in the  
form:
\ba\label{2.7}
[N_{ab},N_{cd}]=(g_{ad}N_{bc}+g_{bc}N_{ad})-(c\leftrightarrow d),
\ea
\ba\label{2.8}
[L_{AB},L_{CD}]=(g_{AD}L_{BC}+g_{BC}L_{AD})-(C\leftrightarrow D),
\ea
\ba\label{2.9}
[N_{ab},L_{CD}]=0,
\ea
where the metric tensor $g_{AB}$ has the following nonzero components:
\ba
g_{AB}=\{g_{ab}, g_{D+1D+1}=-1\}.
\ea
The generators
\ba\label{2.10}
N_{ab}=M_{ab}-{c\over4a^2}Z_{ab}
\ea
form the Lorentz algebra $so(D-1, 1)$ and the generators
\ba\label{2.11}
L_{AB}=\{L_{ab}={c\over4a^2}Z_{ab}, L_{aD+1}=-L_{D+1a}={1\over2a}P_a, 
L_{D+1D+1}=0\}
\ea
form the algebra $so(D-1,2)$\footnote{Note that in the case $D=4$ we obtain the
anti-de Sitter algebra $so(3,2)$.}. The algebra (\ref{2.7})-(\ref{2.9}) is a 
direct sum~$so(D-1,1)\oplus so(D-1,2)$ of the $D$-dimensional Lorentz algebra 
and $D$-dimensional anti-de Sitter algebra, correspondingly.

The quadratic Casimir operators $N_{ab}N^{ab}$, $L_{AB}L^{AB}$ and 
$\e^{abcd}N_{ab}N_{cd}$ of the algebra (\ref{2.7})-(\ref{2.9}) are expressed
in terms of the operators $C_1$ (\ref{2.2}), $C_2$ (\ref{2.3}) and 
$C_3$ (\ref{2.4}) in the following way:
\ba\label{2.12}
N_{ab}N^{ab}-L_{AB}L^{AB}={1\over2a^2}C_1,
\ea
\ba\label{2.13}
N_{ab}N^{ab}={1\over16a^4}C_2,
\ea
\ba\label{2.14}
\e^{abcd}N_{ab}N_{cd}={1\over16a^4}C_3.
\ea

\sect{Supersymmetric $o(N)$ generalization}

In the case $D=4$ dimensions the extended Poincar\'e algebra 
(\ref{2.1}) admits
the following supersymmetric $o(N)$ generalization:
\begin{eqnarray}
\{Q_{\a i},Q_{\b j}\}=-d\left\{\left[{2a\over c}(\g^aC)_{\a\b}P_a
+(\sigma^{ab}C)_{\a\b}Z_{ab}\right]g_{ij}-{4a^2\over c}C_{\a\b}I_{ij}\right\},
\nonumber
\end{eqnarray}
\begin{eqnarray}
[M_{ab},Q_{\a i}]=-(\sigma_{ab}Q_i)_\a,\nonumber
\end{eqnarray}
\begin{eqnarray}
[P_a,Q_{\a i}]=a(\g_aQ_i)_\a,\nonumber
\end{eqnarray}
\begin{eqnarray}
[Z_{ab},Q_{\a i}]=-{4a^2\over c}(\s_{ab}Q_i)_\a,\nonumber
\end{eqnarray}
\begin{eqnarray}
[I_{ij},Q_{\a k}]=Q_{\a i}g_{jk}-Q_{\a j}g_{ik},\nonumber
\end{eqnarray}
\begin{eqnarray}
[I_{ij},I_{kl}]=(I_{il}g_{jk}+I_{jk}g_{il})-(k\leftrightarrow l),\nonumber
\end{eqnarray}
\begin{eqnarray}
[M_{ab},I_{ij}]=0,\nonumber
\end{eqnarray}
\begin{eqnarray}
[P_a,I_{ij}]=0,\nonumber
\end{eqnarray}
\begin{eqnarray}\label{3.1}
[Z_{ab},I_{ij}]=0,
\end{eqnarray}
where $Q_{\a i}$ are the super-translation generators and $I_{ij}$ are 
generators of the $o(N)$ Lie algebra.

Under such a
generalization the Casimir operator (\ref{2.2}) is modified by adding the terms
quadratic in the super-translation generators $Q_{\a i}$ and quadratic in the 
$o(N)$ generators $I_{ij}$
\begin{eqnarray}\label{3.2}
{\tilde C}_1=P^aP_a+cZ^{ab}M_{ba}+2a^2M^{ab}M_{ab}
&-&{c\over2d}Q_{\a i}(C^{-1})^{\a\b}Q_{\b j}g^{ij}+a^2I^{ij}I_{ij}\cr\nn&\stackrel{\rm def}{=}&
X_KH_1^{KL}X_L,
\end{eqnarray}
whereas the form of the rest quadratic Casimir operators (\ref{2.3}) and
(\ref{2.4}) is not changed. In (\ref{3.2}) $X_K=\{P_a, Z_{ab}, M_{ab}, I_{ij},
Q_{\a i}\}$ 
is a set of the generators for the also semi-simple $o(N)$-extended 
superalgebra (\ref{2.1}), (\ref{3.1}).

The tensor
\ba\label{3.3}
H^{KL}=vH_1^{KL}+wH_2^{KL}=(-1)^{p_Kp_L+p_K+p_L}H^{LK}
\ea
is invariant with respect to the adjoint representation
\ba
H^{KL}=(-1)^{(p_K+p_M)(p_L+1)}H^{MN}{U_N}^L{U_M}^K,\no
\ea
where $p_K=p(K)$ is a Grassmann parity of the quantity $K$. 
In (\ref{3.3}) $v$ and $w$ are arbitrary constants and nonzero elements of the
matrix $H_2^{KL}$ equal to the elements of the matrix $H_2^{kl}$ followed from
(\ref{2.3}). Again, by demanding the invariance with respect to the adjoint
representation of the second rank contravariant tensor
$H^{KL}=(-1)^{p_Kp_L+p_K+p_L}H^{LK}$, we come to the structure (\ref{3.3})
(see also the relation (32) in \cite{dss}).

The semi-simple Lie superalgebra (\ref{2.1}) (\ref{3.1}) has the
non-degenerate Cartan-Killing metric tensor $G_{KL}$ (see the relation (A.5)
in the Appendix A) which is invariant with respect to the co-adjoint
representation
\ba
G_{KL}=(-1)^{(p_L+p_N)(p_K+1)}{U_L}^N{U_K}^MG_{MN}.\no
\ea
With the use of the inverse metric tensor $G^{KL}$
\ba
G^{KL}G_{LM}=\d_M^K\no
\ea
we can construct the quadratic Casimir operator (see the relation (A.8) in
the Appendix A) which takes the following expression in terms of the
Casimir operators (\ref{2.3}) and (\ref{3.2}):
\ba\label{3.4}
X_KG^{KL}X_L={1\over4(6-N)a^2}\left({\tilde C}_1-{10-N\over32a^2}C_2\right),
\ea
that meets the particular choice of the constants $v$ and $w$ in (\ref{3.3}).

In the $D=4$ case the extended superalgebra (\ref{2.1}), (\ref{3.1}) can be
rewritten in the form of the relations (\ref{2.7})-(\ref{2.9}) and the 
following ones:
\ba\label{3.5}
\{Q_{\a i},Q_{\b j}\}=-{4a^2d\over c}\left[(\S^{AB}C)_{\a\b}L_{AB}g_{ij}
-C_{\a\b}I_{ij}\right],
\ea
\ba\label{3.6}
[L_{AB},Q_{\a i}]=-(\S_{AB}Q_i)_\a,
\ea
\ba\label{3.7}
[L_{AB},I_{ij}]=0,
\ea
\ba\label{3.8}
[N_{ab},Q_{\a i}]=0,
\ea
\ba\label{3.9}
[N_{ab},I_{ij}]=0,
\ea
where
\ba
\S_{AB}={1\over4}[\G_A,\G_B],\quad\G_A=\{\g_a\g_5,\g_5\},\no
\ea
\ba
\{\g_a,\g_b\}=2g_{ab},\quad g_{ab}=diag(-1,1,1,1),\no
\ea
\ba
\g_5=\g_0\g_1\g_2\g_3.\no
\ea
The generators $N_{ab}$ (\ref{2.10}) form the Lorentz algebra $so(3,1)$ and 
the generators $L_{AB}$ (\ref{2.11}), $I_{ij}$, $Q_\k$ form the 
orthosymplectic algebra
$osp(N,4)$. We see that superalgebra (\ref{2.7})-(\ref{2.9}), (\ref{3.5})-
(\ref{3.9}) is a direct sum $so(3,1)\oplus osp(N,4)$ of the 4-dimensional 
Lorentz algebra and 4-dimensional super-AdS algebra, respectively.

In this case the Casimir operator (\ref{2.12}) is modified as follows:
\ba
N_{ab}N^{ab}-L_{AB}L^{AB}-{c\over4a^2d}Q_\k(C^{-1})^{\k\l}Q_\l+
{1\over2}I^{ij}I_{ij}={1\over2a^2}{\tilde C}_1,\no
\ea
while the form of the quadratic Casimir operators (\ref{2.13}) and (\ref{2.14})
is not changed.

It is remarkable enough that in the particular case $N=10$ the second term in
the right hand side of the relation (\ref{3.4}) vanishes and
\ba
G^{KL}=-{1\over16a^2}H_1^{KL}.\no
\ea
If in this case we consider a gauge group, introduce a gauge form
\ba
A=dx^\m A_\m^K X_K\no
\ea
and find a field strength
\ba
dA+A\wedge A={1\over2}dx^\m\wedge dx^\n F_{\m\n}^KX_K,\no
\ea
then the following Lagrangian:
\ba
L={1\over64a^2}G_{KL} F_{\m\n}^L F_{\r\l}^K g^{\m\r} g^{\n\l} e\no
\ea
can claim for the probable unification of the $N=10$ supergravity with the 
$SO(10)$ GUT model. Here $x^\m$ are space-time coordinates, $e=dete_\m^a$
is a determinant of the tetrad and $g^{\m\n}=e_a^\m g^{ab} e_b^\n$ is a 
metric tensor. The details of this theory will be given elsewhere (see also
\cite{akl}).

There is also another invariant (see, e.g., \cite{dk-gs})
\ba
{\tilde L}=g G_{KL} F_{\m\n}^L F_{\r\l}^K \e^{\m\n\r\l},\no
\ea
where $g$ is the coupling constant.

\sect{Conclusion}

Thus, we proposed the semi-simple second rank tensor $o(N)$-extended 
super-Poincar\'e
algebra in the $D=4$ dimensions. It is very important, since under
construction of the models it is more convenient to deal with the
non-degenerate space-time symmetry. We also constructed the quadratic Casimir
operators for this algebra. The form of these Casimir operators (\ref{2.3}) 
and (\ref{3.2}) for the semi-simple $o(N)$-extended super-Poincar\'e algebra 
indicates that 
the components of an irreducible representation for this algebra are 
distinguished by the mass, angular momentum and quantum numbers corresponding 
to the tensor generator $Z_{ab}$, super-translation generators $Q_{\a i}$ and 
quadratic Casimir operator $I^{ij}I_{ij}$ of the internal algebra $o(N)$.
In that way we have generalized the Regge trajectory concept.
We also discussed the probable unification of the $N=10$ supergravity with 
the $SO(10)$ GUT model.

It is interesting to develop the models based on this extended algebra. The
work in this direction is in progress.

\appendix
\sect{Appendix: Properties of Lie superalgebra}

Permutation relations for the generators $X_K$ of Lie superalgebra are
\ba\label{A.1}
\left[X_K,X_L\right\}\stackrel{\rm def}{=}
X_KX_L-(-1)^{p_Kp_L}X_LX_K={f_{KL}}^MX_M.
\ea
Structure constants ${f_{KL}}^M$ have the Grassmann parity
\ba\label{A.2}
p\left({f_{KL}}^M\right)=p_K+p_L+p_M=0\pmod2,
\ea
following symmetry property:
\ba\label{A.3}
{f_{KL}}^M=-(-1)^{p_Kp_L}{f_{LK}}^M
\ea
and obey the Jacobi identities
\ba\label{A.4}
\sum_{(KLM)}(-1)^{p_Kp_M}{f_{KN}}^P{f_{LM}}^N=0,
\ea
where the symbol $(KLM)$ means a cyclic permutation of the quantities
$K$, $L$ and $M$.
In the relations (\ref{A.1})-(\ref{A.4}) an every index $L$ takes either a
Grassmann-even value $l$ $(p_l=0)$ or a Grassmann-odd one $\l$ $(p_\l=1)$.
The relations (\ref{A.1}) have the following components:
\ba
[X_k,X_l]={f_{kl}}^mX_m,\no
\ea
\ba
\{X_\k,X_\l\}={f_{\k\l}}^mX_m,\no
\ea
\ba
[X_k,X_\l]={f_{k\l}}^\m X_\m.\no
\ea

The Lie superalgebra possesses the Cartan-Killing metric tensor
\ba\label{A.5}
G_{KL}=(-1)^{p_N}{f_{KM}}^N{f_{LN}}^M=(-1)^{p_Kp_L}G_{LK}
=(-1)^{p_K}G_{LK}=(-1)^{p_L}G_{LK},
\ea
which components are
\ba
G_{kl}={f_{km}}^n{f_{ln}}^m-{f_{k\m}}^\n {f_{l\n}}^\m,\no
\ea
\ba
G_{\k\l}={f_{\k\m}}^m{f_{\l m}}^\m-{f_{\k m}}^\m{f_{\l\m}}^m,\no
\ea
\ba
G_{k\l}=0.\no
\ea

As a consequence of the relations (\ref{A.3}) and (\ref{A.4}) the tensor with
low indices
\ba\label{A.6}
f_{KLM}={f_{KL}}^NG_{NM}
\ea
has the following symmetry properties:
\ba\label{A.7}
f_{KLM}=-(-1)^{p_Kp_L}f_{LKM}=-(-1)^{p_Lp_M}f_{KML}.
\ea

For a semi-simple Lie superalgebra the Cartan-Killing metric tensor is 
non-degenerate and therefore there exists an inverse tensor $G^{KL}$
\ba
G_{KL}G^{LM}=\d_K^M.\no
\ea
In this case, as a result of the symmetry properties (\ref{A.7}), the
quantity
\ba\label{A.8}
X_KG^{KL}X_L
\ea
is a Casimir operator
\ba
[X_KG^{KL}X_L,X_M]=0.\no
\ea

\end{document}